\documentclass[aps,preprint,showpacs]{revtex4}
\usepackage{epsfig}

\begin{document}

\title{Understanding penta quark with various quark models}

\author{Jialun Ping$^{a,b,c}$, Di Qing$^d$, Fan Wang$^b$, T. Goldman$^e$}

\affiliation{$^a$Department of Physics, Nanjing Normal University,
Nanjing, 210097, China}
\affiliation{$^b$Center for Theoretical Physics, Nanjing
University, Nanjing, 210093}
\affiliation{$^c$School of Physics and
Micro-electronics, Shandong University, Jinan, 250100, China}
\affiliation{$^d$CERN, Ch-1211 Geneva 23, Switzerland}
\affiliation{$^e$Theoretical Division, LANL, Los Alamos, NM87545, USA}

\begin{abstract}
The pentaquark state recently discovered has been studied with
three quark models which either fit the nucleon spin structure or
the $NN$ scattering. A minimum $\Theta^+$ mass of 1620 MeV is
obtained both for the $\frac{1}{2}^\pm$ state. The mixing of
various color structure configurations, which would reduce the
mass of the penta-quark state, should be taking into account in
the calculation of penta-quark mass.
\end{abstract}

\pacs{12.39.Mk, 13.75.Jz, 12.39.Jh}


\maketitle

\bigskip

Eleven groups \cite{exp} claimed recently that they found a
penta-quark state, now called $\Theta^+$, with mass$\sim$1540 MeV,
and width $\Gamma < 25$ MeV. This state is either identified from
the decay product $nK^+$ or $pK_s$, but up to now no experiment
has identified both. In addition, the NA49 collaboration claimed
that they found the anti-decuplet partner $\Xi^{--}$ of $\Theta^+$
\cite{na49}. The HERA-H1 collaboration claimed that they found the
charm penta-quark $\Theta_c$ \cite{h1}.

These measurements might be contaminated by the normal meson
production due to the kinematical reflection \cite{dzi,zav}. The
NA49 claim has been challenged by another CERN group based on
$\Xi$ spectroscopy data with higher statistics \cite{fis}.  HERA-B
$p$-nucleus reaction data has not found the $\Theta^+$
\cite{herab}. BES $J/\Psi$ decay data analysis has not found the
$\Theta^+$ either \cite{bes}. There are other groups that have not
found the $\Theta^+$ \cite{longo}.

The results of the reanalysis of the $K^+d$ and $K^0_Lp$
scattering data are diverse \cite{arn,kel}. W.R. Gibbs reanalyzed
the $K^+d$ scattering data, taking into account the double
scattering, and found a structure corresponding to a resonance
with width of 0.9$\pm$0.2 MeV, which is either a 1.547$\pm$0.002
GeV $\frac{1}{2}^-$ or a 1.559$\pm$0.003 GeV $\frac{1}{2}^+$ state
and $\frac{1}{2}^-$ is favored \cite{gibbs}. In addition, a very
tiny bump had appeared in 1973 CERN $K^+p\rightarrow pK^0_s\pi^+$
inelastic scattering data \cite{les}.

Theoretical studies based on the chiral soliton quark model played
an important role in triggering the $\Theta^+$ searches
\cite{dia}. In the chiral soliton quark model, the $\Theta^+$ is a
member of the anti-decuplet rotational excitation following the
well established octet and decuplet baryons \cite{pra,kop}. The
QCD background of this model has been debated between different
groups \cite{jw,cohen,pob,diak,man,ell}

Various quark models have been proposed to understand the
$\Theta^+$, mainly aimed at understanding the parity, low mass and
narrow width. In the naive quark model \cite{rgg} the ground state
should be an $S$-wave one and this means $\Theta^+$ should be a
negative parity state. The $S$-wave $uudd\bar{s}$ configuration
will have $S$-wave $KN$ components. However both the $I=0,1$ $KN$
$S$-wave phase shifts are negative in the $\Theta^+$ energy region
\cite{has} and this means that $\Theta^+$ can not be an $S$-wave
$KN$ resonance, except the new analysis of Gibbs \cite{gibbs}. On
the other hand, since the $I=0$ $KN$ $P$-wave $P_{01}$ phase
shifts are positive, there might be resonance in this channel and
this is consistent with the $J^p=\frac{1}{2}^+$ predicted by the
chiral soliton quark model. Bag model gave similar result
\cite{bag,isg}.

Various quark correlations have been proposed such as color Cooper
pair and others \cite{jw,kl,shur} aimed to reconcile the
difference between the quark model predictions and the properties
of $\Theta^+$. It seems to be able to get an even parity
penta-quark ground state with small width but hard to get as low
as 1540 MeV mass \cite{car}. For example, Yu. A. Simonov did a
quantitative calculation based on the Jaffe-Wilczek configuration
\cite{jw} by means of the effective Hamiltonian approach. The
calculated $\Theta^+$ mass is about 400 MeV higher than the
observed 1540 MeV \cite{yua}. Different quark interaction
mechanisms such as quark-gluon, Goldstone boson exchange and
instanton interaction have all been tried to understand the
$\Theta^+$ \cite{jw,srg,koc}. It is fair to say up to now there is
no {\it ab initio} dynamical calculation to obtain a $\Theta^+$ mass
as low as 1540 MeV with the constraint that these model parameters
fit the normal hadron spectroscopy \cite{jm,carl}.

QCD sum rule and lattice QCD both have been used to calculate the
penta-quark, the results are diverse. One Lattice QCD group
reported they have not observed either $I=0$ or $I=1$,
$J^p=\frac{1} {2}^{\pm}$ bound penta-quark state \cite{liu}. Two
groups observed odd and one observed even parity state \cite{lat}.
The pitfalls of these lattice QCD calculations have been discussed
in \cite{cfkk}. The $\Theta^+$ is a resonance state and so its
mass should be a complex number. It might be difficult to detect
such a complex number in an Euclidean Monte Carlo calculation. QCD
sum rule calculations are diverse too. Zhu and Sugiyama {\em et
al.} favor negative parity ones while others favor positive parity
\cite{sum}. Kondo {\em et al.} suggested to remove the reducible
part from the correlation function and after do that they got
positive parity penta-quark.

Our group has done three quark model calculations. The first one
is an application of the Fock space expansion model which we
developed to explain the nucleon spin structure \cite{qcw}. The
naive quark model assumes that the baryon has a pure valence $q^3$
configuration. This is certainly an approximation. One expects
there should be higher Fock components,
\begin{equation}
B = aq^3 + bq^3q\bar{q}+\cdots.
\end{equation}
The nucleon spin structure discovered in polarized lepton-nucleon
deep inelastic scattering shows that there are intrinsic
non-perturbative sea quark components in nucleon and indeed the
nucleon spin structure can be explained by a dynamical model of
nucleon where the ground state has about 15$\%$ $q^3q\bar{q}$
component. This means that even the nucleon ground state might be
a mixing of tri- and penta-quark components. In the $\Theta^+$
mass calculation we assume it is a pure $uudd\bar{s}$ five quark
state but with channel coupling. Our results are listed below:
\begin{table}
\caption{Quark shell model calculations}
\begin{tabular}{ccccc} \hline
       & ~~pure $KN$~~  & $~~~KN+K^*N~~~$ & $~~~KN+K_8N_8~~~$
       & $~~~KN+K^*N+K_8N_8$
       \\ \hline
 $S_{01}$ Parity$=-$ & 2282  & 2157    &  1943   & 1766   \\ \hline
 $P_{01}$ Parity$=+$ &  2357.1  &  2356.3    &  2357.0   &  2336.8
 \\  \hline
\end{tabular}
\end{table}

Here the calculated mass is in units of MeV and $K_8N_8$ means the
$K$ and $N$ are both in color octets but coupled to an overall
color singlet.  The $S$-wave state has strong channel mixing: the
amplitudes of $KN$, $K^*N$, $K_8N_8$, $K^*_8N_8$ are $-0.54$,
$-0.29$, $-0.54$, $-0.29$ respectively. On the other hand the
channel mixing is weak for the $P$-wave state: The amplitudes of
$KN$, $K^*N\frac{1}{2}$, $K^*N\frac{3}{2}$, $K_8N_8$,
$K^*_8N_8\frac{1}{2}$, $K^*_8N_8\frac{3}{2}$ are 0.968, $-0.1$,
$\sim$0, $-0.07$, 0.224, $\sim$0 respectively. Here the
$\frac{1}{2}$ and $\frac{3}{2}$ are the channel spin. The $S$-wave
state definitely has a lower mass than that of the $P$-wave; the
channel coupling plays a vital role in reducing the calculated
$S$-wave $\Theta^+$ mass. In obtaining these results we use the
model parameters which fit the nucleon mass with five quark
components mixed to $q^3$ configuration. Neglecting the five quark
component, this model will give $M(N)=1.2$ GeV, $M(K)=0.8$ GeV.
Therefore it is possible to reduce the penta-quark mass further by
taking into account hepta-quark component and other channels
coupling. We like to point out that our calculation is in fact a
quark shell model calculation. The large scale shell model
calculation of nuclear structure shows that it is possible to
obtain approximate correct energy and wave function of a nuclear
state if the Hilbert space is large enough. This should be true
for quark shell model as well.

Lattice QCD and non-perturbative QCD both show that confinement
might be due to gluon flux tube (or gluon string) formation in a
quark system. The ground state energy of the gluon field in a
$q\bar{q}$ meson and $q^3$ baryon can be approximated by a
potential \cite{yu88}
\begin{eqnarray}
&&V_{q\bar{q}}=-\frac{A_{q\bar{q}}}{r}+\sigma_{q\bar{q}}r+C_{q\bar{q}}
\nonumber\\
&&V_{3q}=-A_{3q}\sum_{i<j}\frac{1}{|\vec{r}_i-\vec{r}_j|}
        +\sigma_{3q}L_{min}+C_{3q}, \label{pot} \\
&&\L_{min}=\sum_i L_i. \nonumber
\end{eqnarray}
$L_i$ is the distance between the quark i and the Y-shaped gluon
junction. $\vec{r}_i$ is the position of quark $i$. The first term
in Eq.(\ref{pot}) is the color Coulomb interaction and the second
term is similar to a linear confinement potential.

Most of constituent quark models use a quadratic or linear
potential to model the quark confinement,
\begin{eqnarray}
V_{conf}(\vec{r}_{ij})=-a\vec{\lambda}_i\cdot\vec{\lambda}_j\vec{r}^n_{ij},
\nonumber\\
\vec{r}_{ij}=\vec{r}_i-\vec{r}_j, ~~~~~~   n=1,2.
\label{conf}
\end{eqnarray}
Here $\lambda^a_i$ ($a=1\cdots 8$) is the color SU(3) group
generator. For a single hadron, $q\bar{q}$ mesons or $q^3$
baryons, such a modelling can be achieved by adjusting the
strength constant $a$ of the confinement potential. The color
factor $\vec{\lambda}_i\cdot\vec{\lambda}_j$ gives rise to a
strength ratio $1/2$ for baryon and meson which is almost the
ratio for the minimum length of the flux tube to the circumference
of the triangle formed by three valence quarks of a baryon.

How to extend the confinement potential to multiquark systems is
an open question. There is a lattice QCD calculation of the
penta-quark potential recently \cite{oki}.  The ground state
energy of the gluon field in a penta-quark with color structure
$qq(\bar{3})\bar{s}(\bar{3})qq(\bar{3})$ can be expressed as
\begin{eqnarray}
V_{5q}=\frac{\alpha_s}{4}\sum_{i<j}
\frac{\vec{\lambda}_i\cdot\vec{\lambda}_j}{|\vec{r}_i-\vec{r}_j|}
+\sigma_{5q} L_{min}+C_{5q}.
\end{eqnarray}
Here $qq(\bar{3})$ means a color anti-triplet $qq$ pair. $L_{min}$
is the minimum length of the color flux to connect the five
quarks.

From general SU(3) color group considerations, there might be
other color structures for a penta-quark:
${q^3}(1){q\bar{s}}(1)$;${q^3}(8){q\bar{s}}(8)$;
${qq}(\bar{3}){qq}(\bar{3})\bar{s}(\bar{3})$, etc. The first one
is a color singlet meson-baryon channel; the second is the hidden
color meson-baryon channel; the third is the color structure used
in the Jaffe-Wilczek model. One guesses the energy of these color
configurations can be expressed in a similar manner as those given
in Eq.(2) and (4). Can the two body confinement interaction Eq.(3)
describe the confinement interaction properly for these color
structures? It is a question needed to be studied further
\cite{ww}. A penta-quark state should be a mixing of these color
structures. Our first model calculation mentioned above shows that
channel coupling reduces the calculated ground state penta-quark
mass. It should be true in general that these different color
channel mixing will reduce the ground state energy.

To do a model calculation for a multi-quark system with the above
multi-body interaction and multi-channel coupling is numerically
quite involved. We have developed a model, called the quark
delocalization, color screening model (QDCSM): 1. We
re-parameterize the confinement potential Eq.(3) to take into
account the effect of channels coupling in multi-quark systems
induced by various color structures, which are not possible for a
$q\bar{q}$ meson and $q^3$ baryon; 2. To take into account the
orbital excitations but keep the numerical calculation simple we
still use quark cluster bases but a delocalized quark orbital wave
function is used to allow the multi-quark system to choose its own
favorable configuration, i.e., to allow the multi-quark system to
vary from the asymptotic hadronic cluster state to a genuine
multi-quark state and all intermediate configurations
\cite{wang1}. This model explains the existing $BB$ interaction
data (bound state deuteron and $NN$, $N\Lambda$, $N\Sigma$
scattering) well with all model parameters fixed by hadron
spectroscopy except for only one additional parameter, the color
screening constant $\mu$ which is determined by the deuteron
properties. More important, {\em it is the unique model, so far,
which explains a long-standing fact: The nuclear force and the
molecular force are similar except for the obvious difference of
length and energy scales; the nucleus is approximately an A
nucleon system rather than a 3A quark system} \cite{wang2}.

\begin{center}
\epsfxsize=4.0in \epsfbox{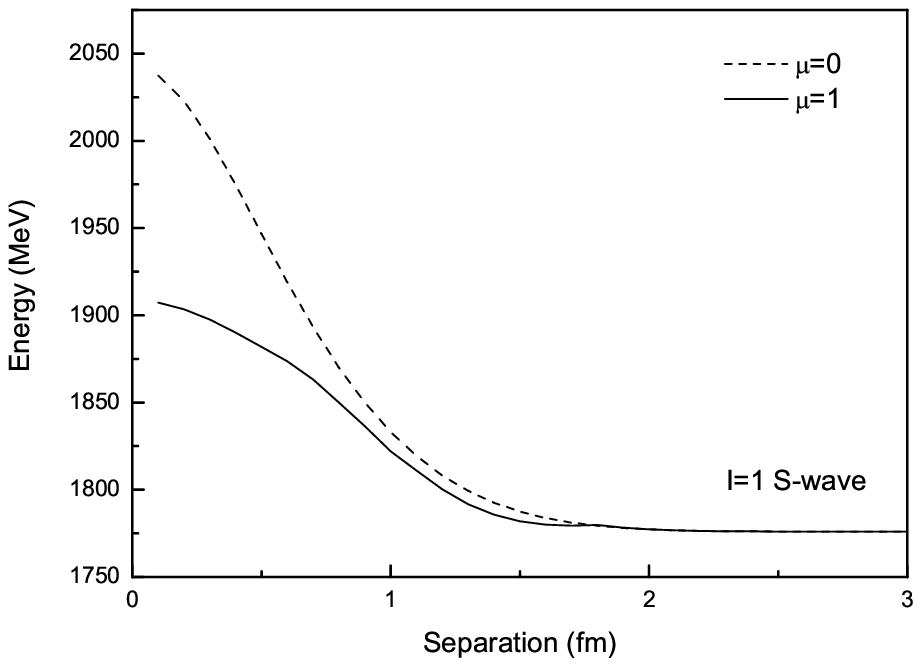}

Fig.1. The effective potentials for I=1 $S$-wave.

\epsfxsize=4.0in \epsfbox{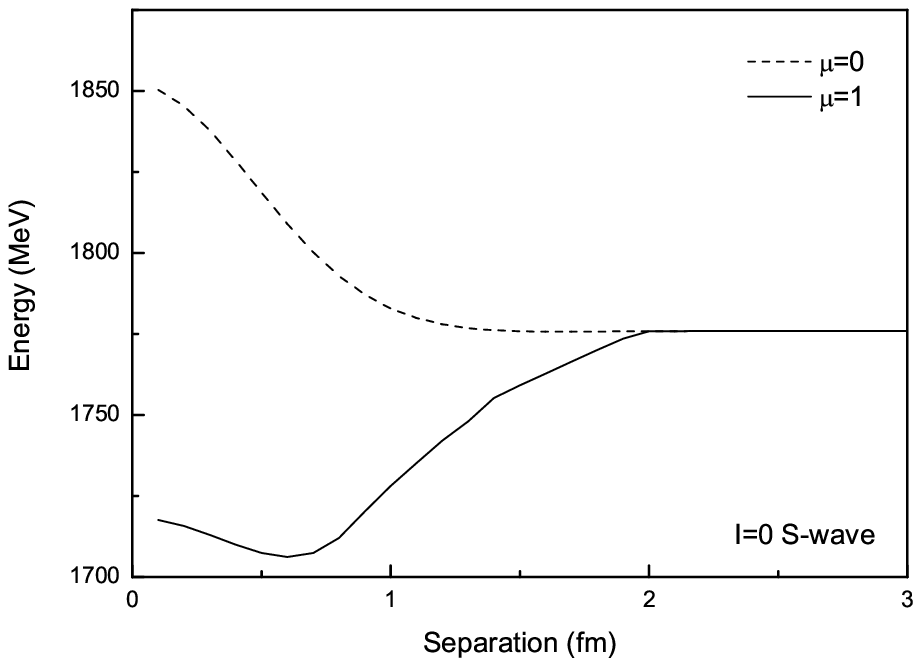}

Fig.2. The effective potentials for I=0 $S$-wave.
\end{center}

The penta-quark mass has been calculated with this model (QDCSM)
in a single color singlet $KN$ channel approximation. As explained
before, the effect of the coupling of other color structures and
orbital excitations is assumed to have been included in the
modelling of QDCSM. In the $I=1$ $S$-wave $KN$ channel, a pure
repulsive effective interaction is obtained as shown in Fig.1
(corresponding to curve with $\mu=1$). This is consistent with the
$KN$ scattering data and helps to rule out the $I=1$ possibility
for the $\Theta^+$. For comparison the naive quark model result is
also shown in Fig.1 (corresponding to curve with $\mu=0$), which
shows a stronger repulsion. In the $I=0$ $S$-wave $KN$ channel, an
effective attraction is obtained and shown in Fig.2 ($\mu=1$
curve). This is inconsistent with the VPI $KN$ scattering phase
shifts \cite{has} but might be consistent with Gibbs new results
\cite{gibbs}. A $\Theta^+$ mass of 1706 MeV is obtained from the
minimum of Fig.2 $\mu=1$ curve. Part of the overestimate of the
$\Theta^+$ mass is due to the overestimate of $K$ mass, which is
650 MeV in this approach. This can be eliminated as follows: One
first gets an effective interaction potential from Fig.2 $\mu=1$
curve by subtracting its asymptotic value, then add a zero point
oscillating energy $\frac{3{\hbar}^2}{4\mu_{KN}R_0^2}$ ($\mu_{KN}$
is the reduced mass of $K$ and $N$, $R_0=0.6 fm$ is the minimum
point of the effective potential) and the rest mass of $N$ and
$K$. In this way one obtains $M(\Theta^+)=1615 MeV$. It is still
about 75 MeV higher than the observed value 1540 MeV. More precise
dynamical calculation might reduce the $\Theta^+$ mass further.
Fig.2 $\mu=0$ curve shows the naive quark model result for
comparison. It is almost a pure repulsive interaction and will not
accommodate a $\Theta^+$ resonance. The $K^+n$ effective potential
is shown in Fig.3, which is a very weak
repulsive interaction resulted from a cancellation of the $I=1,0$
channel ones. This result shows it is hard to get reliable $K^+n$
scattering amplitude from $K^+d$ scattering data because it is a
small component in comparison to the big $K^+p$ amplitude. Gibbs
analysis shows the additional complications \cite{gibbs}. To get
reliable $I=0$ scattering phase shifts from $K^+n$ is even harder
because one has to get two big ones (corresponding to $I=1,0$
separately), which have opposite sign and so cancel each other,
from a small one corresponding to $(I=1)+(I=0)$. For the $P$-wave
channels, we only obtain spin averaged effective $KN$ interactions
because the spin-orbit coupling has not been included yet. In the
$I=0$ channel, there is a strong attraction (shown in Fig.4), as
wanted in other quark models with correlations. However in our
model the $P$-wave attraction is not strong enough to overcome the
kinetic energy increase to reduce it to be a ground state. This is
consistent with the lattice and QCD sum rule results
\cite{lat,sum}. In the $I=1$ channel, only a very weak attraction
is obtained.  This rules out the $I=1$ $\Theta^+$ again.

\begin{center}
\epsfxsize=4.0in \epsfbox{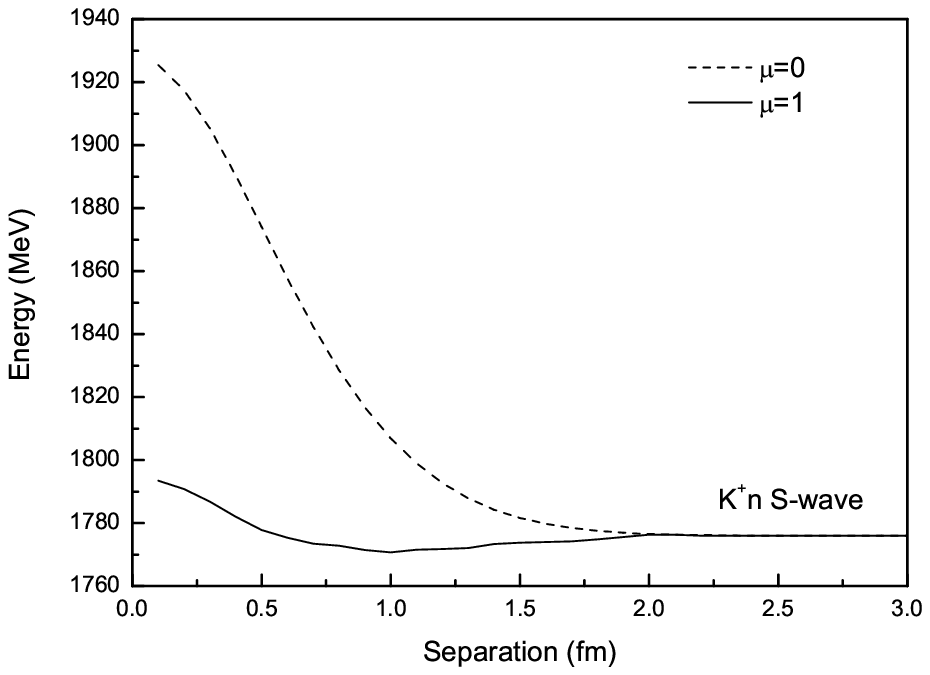}

Fig.3. The effective potentials for $K^+n$ $S$-wave.

\epsfxsize=4.0in \epsfbox{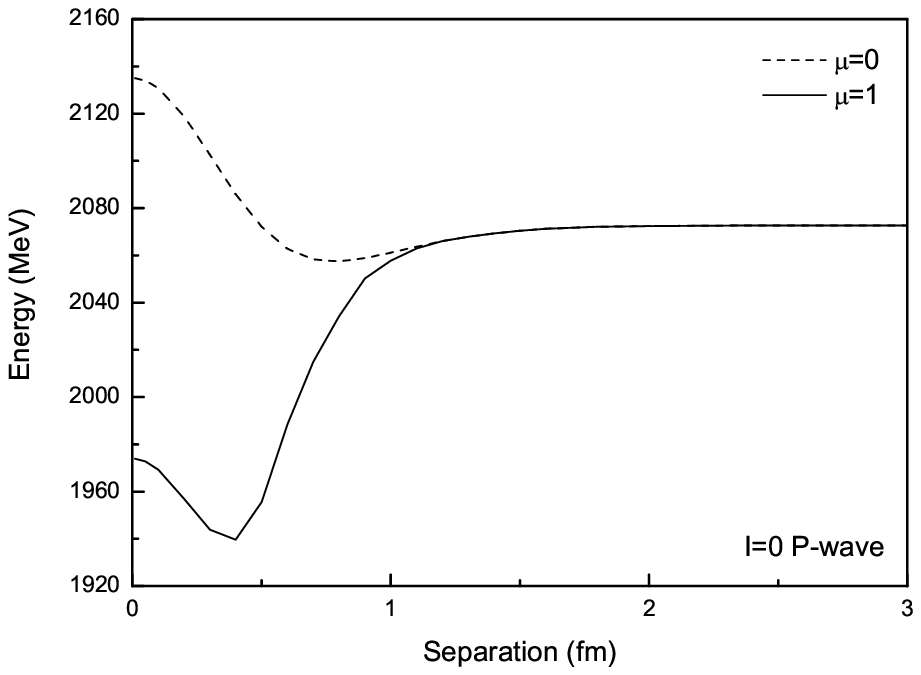}

Fig.4. The effective potentials for I=0 $P$-wave.

\epsfxsize=4.0in \epsfbox{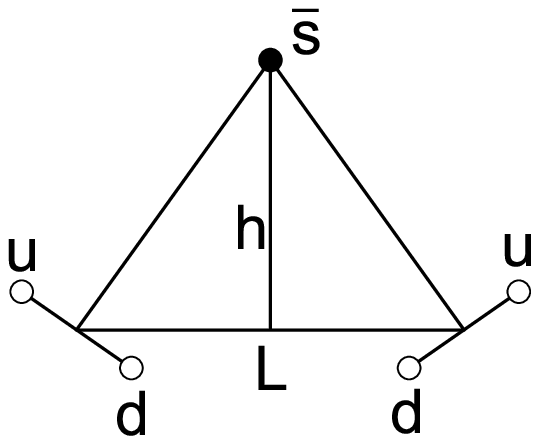}

Fig.5. The Jaffe-Wilczek configuration.

\epsfxsize=4.0in \epsfbox{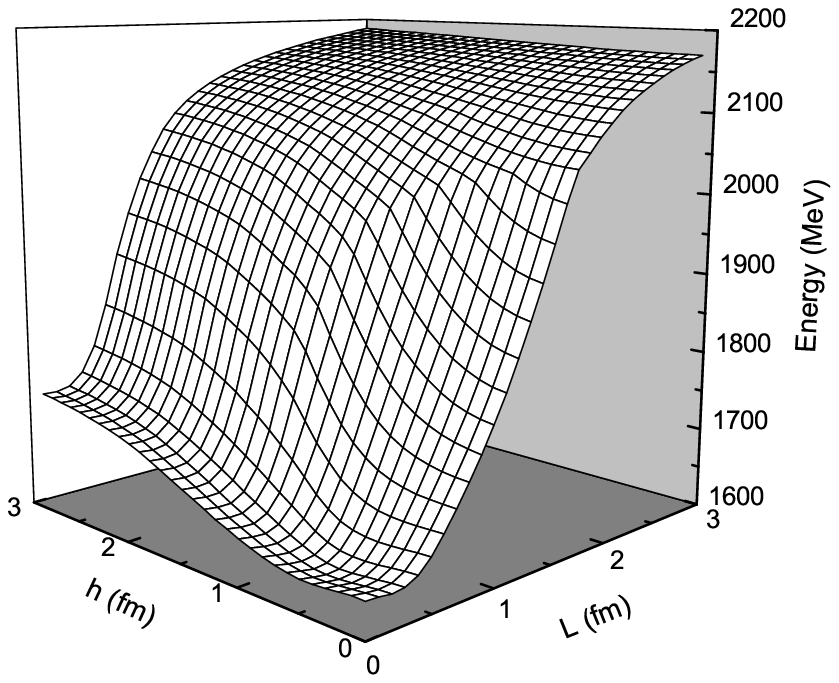}

Fig.6. The energies for Jaffe-Wilczek configuration ($\mu=1$).
\end{center}

In a third model, the Jaffe-Wilczek configuration
$\{ud\}\{ud\}\bar{s}$ is used but the four nonstrange quarks are
totally antisymmetrized. The space part is fixed to be an
equilateral triangle with the two diquarks sitting at the bottom
corners and the $\bar{s}$ at top (see Fig.5). The height $h$ and
the length $L$ of the bottom side of the triangle are taken as
variational parameters in addition to the quark delocalization. A
three body variational calculation with the QDCSM has been done.
The minimum of this variational calculation is 1621 MeV
corresponding to a triangle of $h=0.6 fm, L=0.6 fm$ (see Fig.6).
The $\Theta^+$ mass is similar to our second model one. Fig.7
shows the result obtained with the naive quark model Hamiltonian,
the minimum is 1799 MeV corresponding to a vanishing triangle.

\begin{center}
\epsfxsize=4.0in \epsfbox{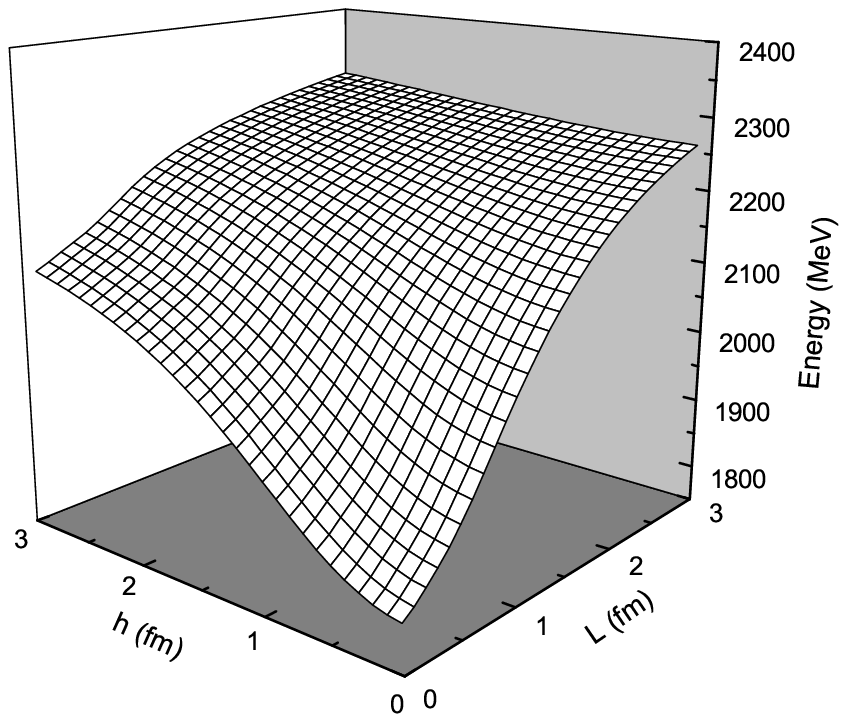}

Fig.7. The energies for Jaffe-Wilczek configuration ($\mu=0$).
\end{center}

Based on these three model results  the $IJ^P=0\frac{1}{2}^\pm$
are both possible to be the ground state of the $\Theta^+$.  A
multi channel coupling quark model calculation with various
nontrivial color structures are needed to check if the observed
$\Theta^+$ mass can be obtained in quark model approach and which
parity is the ground state.

Suppose the $\Theta^+$ is finally verified to be a 1540 MeV narrow
width ($\sim$ 1 MeV) $IJ^P=0\frac{1}{2}^+$ state. Then an
interesting scenario similar to that of nuclear structure at the
1940's turned to the 1950's will recur. The low lying, even parity
rotational excitation of nuclei is hard to explain by the naive
Mayer-Jenson nuclear shell model; Bohr and Mottelson had to
introduce the rotational excitation of a deformed liquid drop
model. Later, nucleon Cooper pairs were introduced because of the
strong short range pairing correlation. In 1970's-1980's, an
$S$-$D$ Cooper pair interaction boson model was developed and the
collective rotation was re-derived from this model which is based
on Mayer-Jenson's nuclear shell model but with nucleon pair
correlation. In the description of the pentaquark, one has
introduced the chiral soliton rotational excitation, quark color
Cooper pairs and much more. The historical lessons of nuclear
structure study might be a good pharos to light the way for the
study of hadron structure.

Summary: Multi-quark states have been studied for about 30 years.
The $\Theta^+$, if further confirmed, will be the first example.
Once the multi-quark "Pandora's box" is opened, the other
multi-quark states: tetra-quark, hexa-quark (or dibaryon), etc.,
can no longer be kept inside. One expects they will be discovered
sooner or later and there are claims that some tetra-quark states
have been observed \cite{babar,cleo,belle,cdf}. A new landscape of
hadron physics will appear and it will not only show new forms of
hadronic matter but will also exhibit new features of low energy
QCD.

Nonperturbative and lattice QCD have revealed the color flux tube
(or string) structure of the $q\bar{q}$, $q^3$ and even
$q^4\bar{q}$ states. The multi-quark system will have more color
structures. How do these color structures interplay within a
multi-quark state? Nuclear structure seems to be understood in
terms of colorless nucleons within a nucleus. Multi-quark states
might be not so. We emphasized that the effect of non-trivial
color structures of multi-quark system should be studied. The low
mass and narrow width of the $\Theta^+$ might be related to such
new structures instead of to residual interactions.

We apologize to those authors whose contributions in this field
have not been mentioned due to the limitation of space. This work
is supported by the NSFC under grant 90103018, 10375030, "333
Project" fund of Jiangsu Province and in part by the US Department
of Energy under contract W-7405-ENG-36.

\end{document}